\documentclass[12pt,preprint]{aastex}
\newcommand{\nraoblurb}{The National Radio Astronomy Observatory is
a facility of the National Science Foundation operated under cooperative
agreement by Associated Universities, Inc.}
\newcommand{\smtoblurb}{The Heinrich Hertz Telescope is operated by the
Submillimeter Telescope Observatory on behalf of Steward Observatory and
the Max-Planck-Institut f{\"u}r Radioastronomie.}
\newcommand{\bimablurb}{Multichannel Image Reconstruction, Image Analysis,
and Display software developed by Berkeley-Illinois-Maryland Association.}
\newcommand{\m}{$\,{\rm m}$}
\newcommand{\ks}{${\,{\rm km\, sec^{-1}}}$}
\newcommand{\msun}{$\,M_\odot$}
\newcommand{\gyr}{$\,{\rm Gyr}$}
\newcommand{\mhz}{$\,{\rm MHz}$}
\newcommand{\ghz}{$\,{\rm GHz}$}

\newcommand{\cma}{$\,{\rm cm^{-3}}$}
\newcommand{\cmb}{$\,{\rm cm^{-2}}$}
\newcommand{\he}[1]{$^{#1}{\rm He}$}
\newcommand{\co}[2]{$^{#1#2}{\rm CO}$}
\newcommand{\car}[2]{$^{#1#2}{\rm C}$}
\newcommand{\ratco}{$^{12}{\rm CO}/^{13}{\rm CO}$}
\newcommand{\ratc}{$^{12}{\rm C}/^{13}{\rm C}$}
\newcommand{\hii}{H~{$\scriptstyle {\rm II}$}}
\newcommand{\hep}[1]{$^{#1}{\rm He}^{+}$}

\newcommand{\her}[1]{$^{#1}{\rm He}/{\rm H}$}
\newcommand{\expo}[1]{${10^{#1}}$}
\newcommand{\nexpo}[2]{${#1 \times 10^{#2}}$}
\newcommand{\ngc}[1]{NGC\thinspace #1}

\begin{document}

\title{CO Isotopes in Planetary Nebulae}


\shortauthors{Balser et al. 2002}

\author{Dana S. Balser\altaffilmark{1}, Joseph
P. McMullin\altaffilmark{2}, \& T. L. Wilson\altaffilmark{3,4}}

\altaffiltext{1}{National Radio Astronomy Observatory,
P.O. Box 2, Green Bank WV 24944}
\altaffiltext{2}{National Radio Astronomy Observatory, 520 Edgemont
Road, Charlottesville, VA 22903}
\altaffiltext{3}{SMTO, Steward Observatory, University of Arizona,
Tucson, AZ 85720}
\altaffiltext{4}{Max-Planck-Institut f{\"u}r Radioastronomie, 
53121 Bonn, Germany}

\begin{abstract}

Standard stellar evolution theory is inconsistent with the observed
isotopic carbon ratio, \ratc, in evolved stars.  This theory is also
inconsistent with the \her3\ abundance ratios observed in Galactic
\hii\ regions, when combined with chemical evolution theory.  These
discrepancies have been attributed to an extra, non-standard mixing
which further processes material during the RGB and should lower both
the \ratc\ and \her3\ abundance ratios for stars with masses $\le
2$\msun.  Measurements of isotopic ratios in planetary nebulae probe
material which escapes the star to be further processed by future
generations of stars.

We have measured the carbon isotopic abundance ratio, \ratc, in 11
planetary nebulae (PNe) by observing the J=2$\rightarrow$1 and
J=3$\rightarrow$2 millimeter transitions of \co12\ and \co13\ in
molecular clouds associated with the PNe.  A large velocity gradient
(LVG) model has been used to determine the physical conditions for
each PNe where both transitions have been detected.  We detect both
\co12\ and \co13\ in 9 PNe.  If \ratco\ = \ratc, the range of \ratc\
is 2.2--31.  Our results support theories which include some form of
extra mixing.

\end{abstract}


\section{Introduction}

Measuring isotopic abundance ratios is one method of probing how stars
of various masses process material and how the Galaxy redistributes
this material over time---that is, stellar and Galactic chemical
evolution.  Some of the light elements H, $^2$H, $^3$He, $^4$He, and
$^7$Li, were produced during the era of primordial nucleosynthesis, so
these ratios provide additional constraints for Big Bang
nucleosynthesis.  \hii\ regions or molecular clouds are examples of
zero-age objects which are young relative to the age of the Galaxy.
Therefore these abundance ratios chronicle the results of billions of
years of Galactic chemical evolution.  Observations of isotopic ratios
in stars and planetary nebulae (PNe) measure locally how stars
influence chemical evolution.  In particular, PNe probe material which
has been ejected from low-mass ($M \le\ 2$\msun) to intermediate-mass
($M \sim 2-5$\msun) stars to be further processed by future
generations.

Measurements of \he3\ in the Galaxy have led to what has been called
``The \he3\ Problem'' \citep{galli97}.  Standard stellar evolution
theory predicts that \he3\ is produced in significant quantities by
stars of $1-2$\msun\ and that PNe with progenitors in that mass range
should have \her3\ $\sim$ several $\times 10^{-4}$
\citep{iben67, rood72, vassiliadis93, weiss96}.  (All abundance ratios
will be by number unless otherwise noted.)  Galactic chemical
evolution models which integrate over a range of initial masses
predict that \he3\ is produced in prodigious amounts \citep{rood76,
dearborn86}.  A survey of 6 PNe produced a detection of \he3\ in
\ngc{3242} \citep{rood92, balser99b} and probable detections in two
other sources \citep{balser97} with \her3\ $\sim$ \expo{-4} --
\expo{-3} as predicted by standard stellar evolution models.  This PNe
sample was highly biased, however, and was selected to maximize the
likelihood of a detection.  Observations of \he3\ in protosolar
material \citep{geiss93}, the local interstellar medium
\citep{gg96}, and Galactic \hii\ regions \citep{balser99} all indicate
a \her3\ abundance ratio of $\sim$ \nexpo{2}{-5}.  Although \hii\
regions are expected to be zero-age objects there is no evidence for
any stellar \he3\ enrichment during the last 4.5\gyr\
\citep{bania02}.  The observed abundances of \he3\ throughout
the Galaxy when combined with the observations that some stars do indeed produce
\he3\ strongly disagree with standard chemical evolution models
\citep{steigman92, flam94, steigman95, dearborn96, prantzos96,
fields96, scully96, galli97}.

Anomalies in the \ratc\ abundance ratio in red giant stars has
stimulated studies of non-standard stellar evolution models
\citep{charbonnel94, charbonnel95, wasserburg95, weiss96,
sackmann99, boothroyd99}.  All of these models include some form of
extra mixing which transports material in the convective envelope to
lower regions which are hot enough for further nuclear processing and
then transports it back up into the convective envelope.  The extra
processing that occurs is called cool bottom processing (CBP).  

During the first dredge-up on the red giant branch (RGB), the
convective envelope reaches regions where \car12\ had previously been
processed into \car13\ and $^{14}$N; thus we expect \ratc\ to decrease
from $\sim$ 90 to 25 for both low-mass and intermediate-mass stars
\citep{iben64, iben84}.  Observations of carbon in the surfaces of red
giant stars reveal lower \ratc\ ratios (5--20) for stars below 2\msun\
than theoretically expected (e.g., Gilroy 1989). Studies which measure
the evolution of the surface \ratc\ as a star ascends the RGB in M67
\citep{gilroy91} and in field stars \citep{charbonnel98} indicate that
additional processing occurs between the base and tip of the RGB. This
has been attributed to extra mixing below the convective envelope.

During the asymptotic giant branch (AGB) the theoretical abundances
depend on the interplay between dredge-up events, mass loss, hot
bottom burning, and non-standard mixing (CBP).  During the third
dredge-up on the AGB significant pollution of \car12\ will increase
the \ratc\ ratio if hot bottom burning does not take place
\citep{forestini97}.  In low-mass stars, \ratc\ may decrease to
$\sim 4$ if AGB cool bottom processing occurs
\citep{wasserburg95}. For AGB stars more massive than $\sim 4$\msun,
hot bottom burning drives the \ratc\ towards its equilibrium value of
3.5 \citep{frost98}.

Observations of abundances in PNe measure the cumulative effects of
different mixing and nuclear processing events during the entire
evolution of their progenitors.  In general the abundance of \he3\
should follow the \ratc\ ratio; that is, if \he3\ is destroyed then
the \ratc\ ratio should decrease.  Therefore given the current
observational challenges of detecting \he3\ in PNe, an alternative
approach is to measure \ratc.

There have been several studies of \ratc\ in the molecular gas in PNe
using millimeter transitions of CO \citep{bachiller89, jaminet91,
cox92, bachiller97, palla00}. In these studies it is assumed that the
\ratc\ ratios measured in molecular clouds near the PN refer
to the abundances in the PN itself.  Alternatively, Clegg et
al. (1997) and Miskey et al. (2000) have used the \car13\
$\,^{3}_{1/2}{\rm P}^{o}_{0}\, -\, ^{1}_{1/2}{\rm S}_{0}$ transition
at 1909.597\AA\ to measure the \ratc\ ratio in several PNe. From these
studies the measured \ratc\ abundance ratios are between $\sim$ 10 to
40.  Palla et al. (2000; hereafter PBSTG) concluded that in the
majority of PNe the \ratc\ ratio is lower than expected from standard
stellar models and are consistent with non-standard mixing models
which incorporate CBP.

In this paper we present new observations of \co12\ and \co13\ which
enlarge the PN sample.  Usually isotope ratios are estimated using
optically thin lines, however, since the CO intensities are small, and
since the $^{16}$O/$^{18}$O ratios may vary, the \co12\ and \co13\
lines are used to deduce the \ratc\ ratio.  Because of the resulting
uncertainty in opacity, it is important that at least two transitions
of CO be observed to constrain radiative models used to calculate the
\ratc\ abundance ratio.  The observations are discussed in \S{2} and
the general results are in \S{3}.  In \S{4} the \ratc\ abundance
ratios are calculated using the assumption that the optical depths of
both \co13\ and \co12\ are small and that the transitions are
themalized.  More sophisticated methods are explored in \S{5}.  In
\S{6} the implications with respect to stellar and Galactic evolution
are discussed.  A summary is given in \S{7}.

\section{Observations}

The observed sample of PNe are listed in Table 1.  The PNe were
selected from the CO survey of \citet{huggins89} based on the
estimated intensity of \co13\ for an expected \ratco\ ratio of 25.
Listed are the source name, the Galactic coordinates rounded to the
nearest degree, the equatorial coordinates, the observed offset
position, the LSR velocity, the angular size, the progenitor main
sequence mass, and the morphology.  The progenitor masses, $M_{\rm
ms}$, are taken from PBSTG for objects in common or determined by
first using the central star masses determined by \citet{gorny97} or
Stasi\'{n}ska, G\'{o}rny, \& Tylenda (1997) and the initial-final mass
relation of \citet{herwig96}.  The PNe morphology is defined as round
(R), elliptical (E), or bipolar (B) \citep{balick87, manchado96,
huggins96}.

An observing log is summarized in Table 2 which includes the spectral
transitions, the telescope's half-power beam-width, and the velocity
resolution.  Observations of the \co12\ and \co13 (J=2$\rightarrow$1)
rotational transitions were made with the National Radio Astronomy
Observatory\footnote{\nraoblurb} 12\m\ telescope on 1997 May and 1997
December.  The 200--300\ghz\ dual-polarization SIS receiver was
employed.  Both the filter-bank and hybrid spectrometers were used
simultaneously and were configured to provide spectral resolutions of
1\mhz\ and 2\mhz\ per channel which produce a velocity resolution of
1.4\ks\ and 2.7\ks\ per channel, respectively.  The spectra were
obtained by beam switching plus position switching (BSP).  The
reference position (OFF) was typically 4 arcmin from the source
position (ON).  The pointing should be accurate to within $\sim 5$
arcsec.  The data were calibrated on-line to the radiation temperature
(T$^*_{\rm R}$) scale which corrects for atmospheric attenuation,
radiative loss, and rearward and forward scattering and spillover
\citep{kutner81}.  In this paper we report all intensities in units of
the main beam brightness temperature, T$_{\rm mb}$, using an
efficiency of $\eta^{*}_{\rm m} = 0.5$ which converts the radiation
temperature to the main beam brightness temperature (Mangum 1999).

Observations of the \co12\ and \co13\ (J=3$\rightarrow$2) transitions
were made with the Henrich Hertz Telescope (HHT)\footnote{\smtoblurb}
during February 1998 and May 1998.  The facility 345\ghz\
dual-polarization SIS receiver was employed, although because of
technical difficulties only one channel was used. The spectrometers
consisted of a 1\ghz, 1024-channel AOS and a 250\mhz, 1024-channel AOS
with spectral resolutions of 1.0\mhz\ (0.87\ks) and 0.40\mhz\
(0.35\ks) per channel, respectively.  Beam switching was used to
produce the spectra using a procedure called WSWITCH with a reference
offset from the source by 4 arcmin. (This procedure is effectively
identical to BSP at the 12\m.)  The pointing model was determined to
be accurate to within 1--2 arcsec for most locations in the sky.  The
data were calibrated on-line to the T$^{\prime}_{\rm A}$ temperature
scale which corrects for atmospheric attenuation.  The line
intensities were scaled to main beam brightness temperatures using
several standard calibration sources ($\chi$ Cyg, R Aql, and
IRC+10216).  These intensities are consistent with an assumed
efficiency of $\eta_{\rm mb} = 0.55$, if the mixer sideband ratios are
equal.  We estimate that the uncertainties in the intensity scale are
between 15--20\% \citep{wilson01}.

\section{Results}

The results are summarized in Table 3.  Listed are the source name,
the transition, the peak intensity ($T_{\rm peak}$), the
root-mean-squared ($RMS$) variations in the spectral baseline, the
velocity of the centroid ($V_{\rm centroid}$), the full-width
half-maximum of the linewidth based on a single-component Gaussian fit
($\Delta{V}$), and the integrated intensity ($I$).  Only $\Delta{V}$
is based on a Gaussian model.  The data were analyzed independently
using the spectral line analysis software CLASS, UniPops, and COMB.
The majority of the sources were detected in both the \co12\ and
\co13\ J=2$\rightarrow$1 transitions. Four objects were also detected
in the J=3$\rightarrow$2 transitions which require higher excitation.

Figure 1 summarizes the spectra taken with both telescopes.  The first
two columns correspond to the NRAO 12\m\ J=2$\rightarrow$1 transitions
while the last two columns correspond to the HHT 10\m\
J=3$\rightarrow$2 transitions.  The spectra have been shifted in
velocity to align the profiles relative to the LSR velocity in Table
1.  For several sources there is evidence for complicated velocity
structure (e.g. multiple peaks, asymmetries, etc.).  In these cases
this structure is being averaged over the telescope's beam and thus
our analysis refers to the bulk properties of the molecular emitting
gas.

For PNe observed at both transitions we measure the J=3$\rightarrow$2
transition to be significantly more intense than the J=2$\rightarrow$1
transition. This effect is expected based on solely the radiative
transfer in gas with physical conditions characteristic of PNe.

In Figure 2, the radiation temperature is plotted versus the upper J
level rotational transition for CO over a range of densities. Across a
broad range of physical conditions, the J=3$\rightarrow$2 transition
produces the brightest intensity; only in the lowest volume
density-column density models is the J=2$\rightarrow$1 intensity
higher. The generally better coupling between smaller beams (obtained
at higher frequencies) and PN structure also suggests that CO searches
in PNe may be better carried out in the sub-millimeter regime;
atmospheric and hardware limitations may mitigate this effect.

\section{\ratc\ Abundance Ratios}

The derived \ratc\ ratios for our PN sample are listed in Table 4.  In
columns 3 and 4 the \ratc\ ratios are shown for the J=2$\rightarrow$1
and J=3$\rightarrow$2 transitions, respectively.  Values based on the
integrated intensities are expected to be more accurate, but values
based on the peak intensities are also given (in parentheses).  The
\ratco\ abundance ratios are believed to be a direct measure of the
\ratc\ ratio; therefore we have assumed that \ratc\ = \ratco.
Chemical fractionation will lower the observed \ratco\ ratios, but
this effect should be reduced by the observed high temperatures in the
molecular gas ($\sim 25-60$ K, Bachiller et al. 1997).  Selective
photo-dissociation of \co13\ causes an increase in the \ratco\ ratio
but this will be counterbalanced by the charge exchange reaction
$^{13}$C$^{+}$ + $^{12}$CO $\rightarrow$ $^{12}$C$^{+}$ + $^{13}$CO
\citep{likkel88}.

The range of \ratc\ ratios are lower than the values predicted by
standard stellar models.  About half of the objects observed in this
study were also included in the sample of PBSTG.  In most cases our
ratios agree with the PBSTG values shown in the last column in Table
4.  The peak main brightness temperatures for our observations are
lower in all cases, as expected for the larger beam in our study.
However the intensity is not proportional to the beam areas.  

These \ratc\ ratios were calculated under the assumptions that the
filling fraction for the gas is the same for both isotopomers, both
rotational transitions are thermalized to the same temperature, and
that the emission lines are optically thin.  The first assumption
seems reasonable based on the spectra shown in Figure 1. A comparison
of the different transitions reveals nearly identical $V_{\rm LSR}$
and line shape characteristics, indicating that similar volumes of gas
are sampled by the two transitions. The excitation characteristics of
the low-J transitions of CO also supports the assumption of a
thermalized, uniform temperature.  The small dipole moment of CO
results in a low critical density for rotational excitation; thus most
PNe have densities high enough for significant CO excitation.  The
effect of line opacity in CO may be accounted for in multi-transition
models of radiation transport.

\section{Large Velocity Gradient Models}

In order to better interpret our results, the radiative transfer in
the PNe is examined through the use of the Sobelev or Large Velocity
Gradient (LVG) approximation \citep{sobolev58, sobolev60}.  This model
provides a numerical solution to the coupled equations of radiative
transfer and statistical equilibrium for the rotational transitions of
CO (see e.g. Leung \& Liszt 1976).  In particular, we used the RAD package
which is a part of the MIRIAD\footnote{\bimablurb} software package
\citep{mundy92}.  Only the four PNe for which we have
measured multiple transitions of \co12\ and \co13\ emission are
considered.  For each source a two-dimensional grid composed of volume
density ($n$) and column density ($N$), over a range of physical
conditions relevant to molecular gas in PNe, is generated.  We adopt a
kinetic temperature $T_{\rm k} = 60$ K.  This value is towards the
high end of temperatures suggested by \citet{bachiller97}; however,
the higher J transitions of CO typically trace warmer regions and the
LVG grids are consistent with $T_{\rm k} = 60$ K.  Data taken with the
HHT (J=3$\rightarrow$2) have been spatially smoothed to match the
resolution of the 12\m\ (J=2$\rightarrow$1) data.  We assume that the
emission fills the beam for both pairs of transitions, or has a
similar beam filling fraction.  We obtained $\chi^2$ values from a
comparison of the data to the model grids as a measure of the quality
of the fits.

The derived \ratc\ ratios are listed in column 5 of Table 4.  The
quoted uncertainties were determined by propagating the measured error
in the integrated intensities from Table 3.  These statistical
uncertainties may be much smaller than the systematic uncertainties.
For the PNe for which there only exists J=2$\rightarrow$1 detections
the integrated intensity ratios are adopted.  Where the line shape is
parabolic the line is optically thick so the \ratc\ ratios are cited
as lower limits.  For PNe where we have detected both the
J=2$\rightarrow$1 and J=3$\rightarrow$2 transitions an LVG analysis is
performed and a \ratc\ ratio determined.  Below we discuss each of
these sources in more detail.

{\it AFGL 618}---Using \co13\ we derive $n$ = \nexpo{6.8}{4}\cma and
$N$ = \nexpo{6.8}{15}\cmb.  Using these values, we obtain $N$ =
\nexpo{3.2}{16}\cmb\ for \co12, yielding a ratio of \ratco\ $\sim 5$.
Overall this value is consistent with the observations.  This model
overproduces the J=2$\rightarrow$1 emission by about 20\% and the best
fit density to the J=2$\rightarrow$1 is a factor of $\sim$2 lower than
that derived from the \co13.  This may suggest that the observed
\ratco\ ratio is a lower limit, since the \co12\ and \co13\ emission
regions are not identical, or that density, temperature, and abundance
gradients exist which are not isolated in velocity space. Although the
derived opacity in the J=3$\rightarrow$2 and J=2$\rightarrow$1 lines
from the LVG grids is low, the line shape is strongly suggestive of
high opacity. We therefore cautiously cite the adopted ratio as a
lower limit only.

{\it M1-16}---The model for M1-16 produces $n$ = \nexpo{6.8}{4}\cma\
and $N$ = \nexpo{6.8}{15}\cmb\ for \co13, and $N$ =
\nexpo{1.5}{16}\cmb\ for \co12, yielding \ratco\ = 2.2, very similar
to that derived from both the peak and integrated
intensities. Although this value is very low, the model suggests that
the opacity remains low for both transitions (e.g., $\tau < 0.05$ for
\co12).  PBSTG consider this source to be a lower limit, however,
presumably because of high opacity.  Examining the shape of the
spectrum, the profile seems most like the classical flat topped
structure of an optically thin, unresolved shell.  Therefore we adopt
our model-derived value for the ratio in this source.

{\it IRAS 21282+5050}---Using \co13\ we obtain $n$ =
\nexpo{3.2}{5}\cma\ and $N$ = \nexpo{1.0}{15}\cmb. 
This produces a column density of \nexpo{3.2}{16}\cmb\ and a ratio of
\ratco\ = 32.  The volume density for IRAS 21282+5050 is larger than
that obtained from the other sources, perhaps indicative of its
youth. The shape of its spectral profile is difficult to discern given
the signal-noise in our data.  For \co12\ in the J=3$\rightarrow$2
line the opacity should be higher but this line seems more parabolic
than flat topped.  Moreover, the values obtained from the ratios of
the integrated intensities and peak intensities are not consistent,
most likely due to the low signal-to-noise ratio in the \co13\
data. Due to these concerns, we adopt \ratco\ = 32 as a lower limit.

{\it \ngc{7293}}--- Numerous models were run for \ngc{7293} without
obtaining any reasonable fits to the individual \co12\ or \co13\ data.
In addition, the derived physical characteristics for \co13\ were
highly inconsistent with those from \co12.  Since \ngc{7293} is an
evolved PN, the simplest explanation for these results is that
structure in the CO line emitting gas couples differently in the two
beams, so the J=2$\rightarrow$1 and J=3$\rightarrow$2 regions are
distinct and different.  The emission profiles from the lines do not
suggest a high opacity, however.  Since the J=3$\rightarrow$2 \co13\
transition only has a signal-to-noise ratio of $\sim 4$ we adopt the
J=2$\rightarrow$1 ratio.

\section{Discussion}

The carbon isotopic ratio, \ratc, along with other elements are
processed during different stages of stellar evolution such as the red
giant branch (RGB), the asymptotic giant branch (AGB), and the
planetary nebula (PN).  Observations have been made during these
different stages of evolution which can be used to constrain stellar
evolution theory.

Observations of anomalous \ratc\ abundance ratios in RGB field stars
\citep{tomkin75, tomkin76} and open clusters \citep{gilroy89} have
stimulated theoretical studies which introduce an additional mixing
process to reduce \ratc\ in stars \citep{sweigart79, charbonnel94,
charbonnel95, wasserburg95}.  A detailed study of M67 revealed that
the anomalously low \ratc\ ratios occurred only for stars at the tip
of the RGB \citep{gilroy91}.  Reanalysis of 191 field and cluster
giants stars using HIPPARCOS parallaxes has determined that 96\% of
evolved stars, objects which have passed the luminosity bump, have
\ratc\ ratios lower than standard models predict
\citep{charbonnel98b}.  \citet{gratton00} have measured \ratc\ in 62
field metal-poor stars, including 43 additional stars from the
literature.  They determine two distinct mixing episodes.  The first
dredge-up which produces \ratc\ ratios consistent with standard models
(\ratc\ = 20--30); and a second mixing episode occurring after the
luminosity bump, when the molecular weight barrier is removed, where
\ratc\ is reduced to 6--10.  These results are consistent with the
theoretical models which predict extra mixing
\citep{charbonnel94, charbonnel95}.

Additional processing of carbon and other elements occurs on the AGB.
Observations have been made of \ratc\ in AGB stars using mm-wave,
optical, and IR spectroscopy.  CO mm-wave observations of the
circumstellar envelops of carbon and oxygen-rich AGB stars yield
\ratc\ values between $\sim 10-50$ \citep{wannier87, kahane94,
greaves97}.  Evidence for some deep mixing process is suggested from
an optical study of carbon stars where Li-rich objects typically have
low ($< 15$) \ratc\ ratios \citep{abia97}.  IR spectroscopy of cool
carbon stars produces \ratc\ ratios between $30-70$ with a few very
low ratios, \ratc\ $\sim 4$ \citep{lambert86}.  The high \ratc\ values
are expected from third dredge-up of \car12, while depending on the
mass of the star the very low \ratc\ ratios could be explained by hot
bottom burning or by deep mixing.  In contrast \citet{ohnaka96}
determine \ratc\ ratios that are 2--3 times smaller than Lambert et
al. even though there is a significant overlap in the stars observed.
There have been heated discussions about the analysis of spectra using
the iso-intensity method \citep{laverny98, ohnaka98, laverny99,
ohnaka00}.  Nevertheless, it appears that at least some objects do
have reduced \ratc\ ratios compared to that expected from standard
models and that some extra mixing mechanism is at work.

Observations of \ratc\ in PNe measure how carbon has been processed
through both the RGB and AGB phases.  Because this processed material
escapes into the interstellar medium these \ratc\ ratios can be used
to constrain Galactic chemical evolution models.  Figure 3 summarizes
the current status of \ratc\ abundance ratios in PNe.  Plotted are the
determined \ratc\ ratios versus the estimated main sequence progenitor
mass ($M_{\rm ms}$).  Only PNe for which progenitor masses could be
found are shown.  The results from other CO observations are taken
from PBSTG and displayed as triangles.  Measurements using the CIII]
transition are shown as stars.  Notice that there are five PNe that we
have in common with PBSTG.  These objects are indicated by open
triangles.

For clarity only the (statistical) uncertainties in the \ratc\ ratio
are shown in Figure 3.  The uncertainties in the derived progenitor
masses are significant and are mostly due to estimates of the PN
distance and calculations of the progenitor mass from the current
central star mass using some initial mass--final mass relationship.
These uncertainties can be as large as 50\% (e.g, Galli et al. 1997;
PBSTG).  Therefore care must be used when interpreting these results.
The error bars in Figure 3 for \ratc\ ratios are based on the
uncertainties in the measured integrated intensity for each
transition.  Note that the systematic errors may be much larger
than these, in some cases.

Our results are consistent with PBSTG with the exception of \ngc{6720}
where our \ratc\ ratio is a factor of two smaller.  However, the
signal-to-noise ratio for the \co13\ J=2$\rightarrow$1 transition is
only 5 and we do not have multiple transitions for this PN.  Therefore
the \co12\ transition may be optically thick and our \ratc\ value a
lower limit.  However from the line shapes these data are consistent
with optically thin emission.

The \ratc\ ratios range between 2.2--31 with progenitor masses from
$\sim 1-4$\msun.  This indicates that at least some stars do undergo
some non-standard processing such as cool bottom processing (CBP).
Because of the large uncertainties in determining the progenitor mass
it is difficult to make any strong statements about the relationship
of \ratc\ with $M_{\rm ms}$.  Since CBP is expected to operate only
for stars with $M_{\rm ms} < 2$\msun\ the \ratc\ ratios should be
larger for stars above this mass.  Above 4\msun\ hot bottom burning
can drive the \ratc\ ratio towards it equilibrium value of 3.5
\citep{frost98}.  The low \ratc\ values between 2--4\msun\ might
indicate further processing on the AGB or that we have underestimated
the \ratc\ ratios.  It is interesting to note that the lowest \ratc\
values are classified as bipolar PNe.  Bipolar PNe tend to lie closer
to the Galactic plane with a wider distribution of central star masses
and have higher N/O ratios \citep{gorny97}

Since most of the PNe in our sample have only been detected in one
transition we do not have sufficient constraints to determine the CO
line opacity.  If the \co12\ line is optically thick then the \ratc\
values will be underestimated and the points in Figure 3 must be
increased.  Clearly more \ratc\ ratios are required to improve the
statistics, together with multiple CO transitions to constrain
radiative transfer models.  Because of the higher kinetic temperatures
in PNe, observations in the sub-mm regime may be more useful for this
purpose.

A significant effort has been made to detect CO toward the six PNe in
which \he3\ have been measured to directly establish the relationship
between non-standard mixing in low-mass stars and the \he3\ and \ratc\
abundances (PBSTG).  The stellar models predict that non-standard
mixing (CBP) will lower both \her3\ and \ratc.  Based on chemical
evolution models more than 90\% of low-mass stars must undergo CBP to
reconcile the observations of \he3\ in \hii\ regions
\citep{charbonnel98}.  Only one object with a measured \her3\ ratio,
\ngc{6720}, has been detected in CO.  Since the measured \he3\ upper
limit of \her3\ = \nexpo{0.54 \pm\ 0.29}{-3} is consistent with both
the standard model and with models which include CBP, the current data
do not provide any interesting constraints.  Note that the progenitor
mass of \ngc{6720} is slightly above 2\msun\ which is where CBP should
turn off.

CO is typically detected in young PNe where the nebula has not evolved
sufficiently to dissociate the molecular material.  In contrast,
\he3\ is best detected in evolved, older PNe which tend to fill the
telescope's beam.  Therefore it is not a coincidence that there is
little overlap between the two samples.  As pointed out by PBSTG,
observations using the CIII] transition toward PNe will have a better
opportunity to match the \he3\ samples.  Currently, the number of PNe
with either CIII] or \he3\ detections is sparse.  The Green Bank
Telescope will offer a significant improvement in our ability to
detect \hep3\ in PNe and thus increase the current sample of one well
determined ratio (\ngc{3242}) and two probable detections.
Observations of CO in PNe will complement any CIII] data.  Because of
the large uncertainties in estimating the progenitor masses we feel
that the number of detections must be significantly increased to build
a strong observational case for non-standard mixing in PNe.  It is
interesting to note that even if CBP occurs in most low-mass stars it
will not be a significant factor in the chemical evolution of \ratc\
in the Galaxy since the production of carbon is dominated by
intermediate-mass stars (PBSTG).  Therefore, measurements of \he3\ is
one of the few methods to probe how low-mass stars influence the
chemical evolution of the Galaxy.

\section{ Summary}

The following are the main results of the J=2$\rightarrow$1 and
J=3$\rightarrow$2 millimeter transitions of \co12\ and \co13 in 11
Galactic planetary nebulae (PNe).

1. We have detected both \co12\ and \co13\ emission in 9 PNe for the
J=2$\rightarrow$1 transition using the NRAO 12\m\ telescope and in 4
PNe for the J=3$\rightarrow$2 transition using the HHT 10\m\
telescope.  Assuming \ratc\ = \ratco\ the ratios range from 2.2--31.
These results support an extra mixing process in {\it at least} some stars.

2. Large velocity gradient (LVG) models have been applied to the 4 PNe
where both transitions have been detected.  These models were primarily
used to determine if the \co12\ line is optically thick, an effect
which would overestimate the \ratc\ ratio.  A very low \ratc\ ratio of
2.2 was determined for M1-16 which does {\it not} appear to be a lower limit.

3. The J=3$\rightarrow$2 transition is significantly more intense than
the J=2$\rightarrow$1 transition.  This result has been confirmed by
models using typical densities measured in PNe.  This suggests that CO
observations in PN may be better carried out in the sub-millimeter
regime.

\acknowledgements

We thank the technical support staff at both the NRAO 12\m\ telescope
and the SMTO.  In particular, we thank Harold Butner for help during
our observations at the HHT.  We acknowledge fruitful discussions with
Corinne Charbonnel and Bob Rood about stellar evolution.  Lastly, we
thank Daniele Galli for communicating his results before publication.

\newpage

\begin{figure}
\plottwo{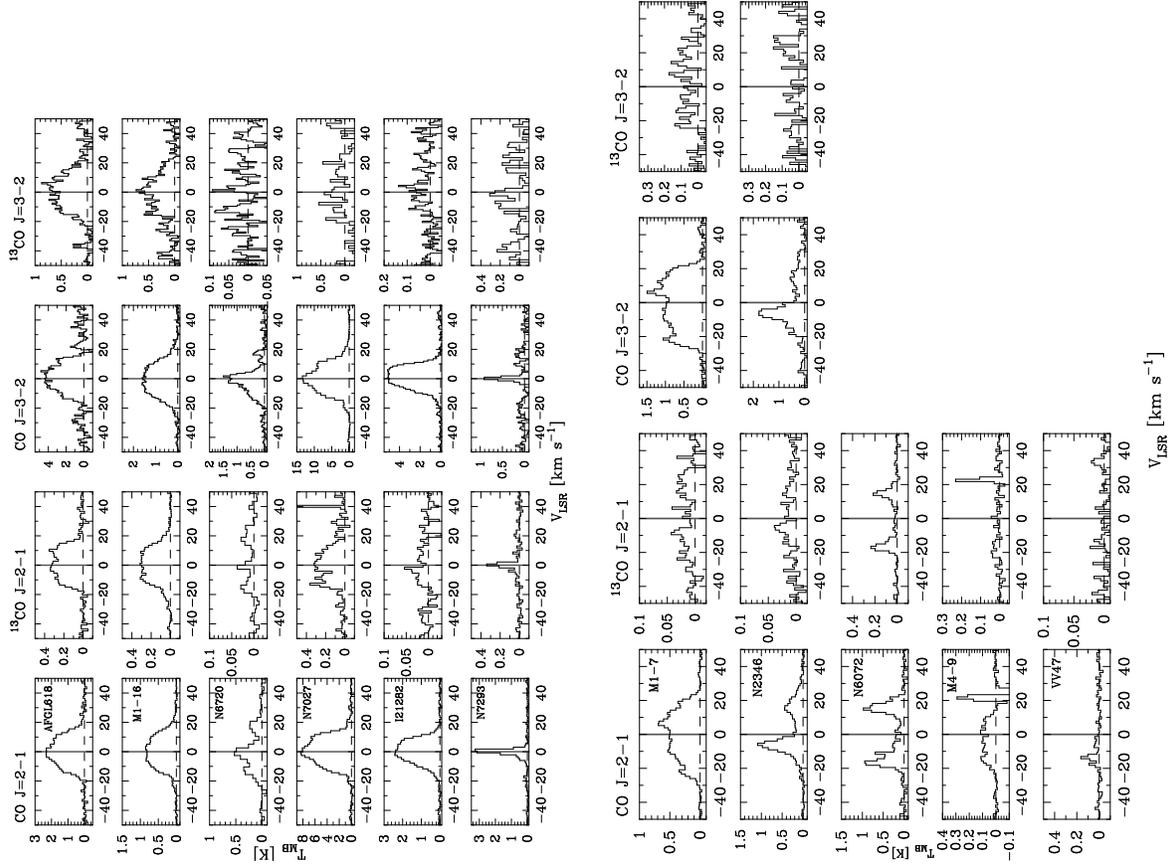}{f1b.eps}
\figcaption[fig1.ps]{NRAO 12\m\ spectra of the $^{12}$CO and $^{13}$CO emission in the
J=2$\rightarrow$1 transition (first two columns) and SMTO HHT spectra
of the $^{12}$CO and $^{13}$CO emission in the J=3$\rightarrow$2
transition (last two columns) for our sample of PNe.  The intensity
scale has been converted into main beam brightness temperature in
units of Kelvin.  The vertical grey line marks the $V_{\rm LSR}$ of
the source.
\label{fig1}}
\end{figure}

\begin{figure}
\epsscale{0.7}
\plotone{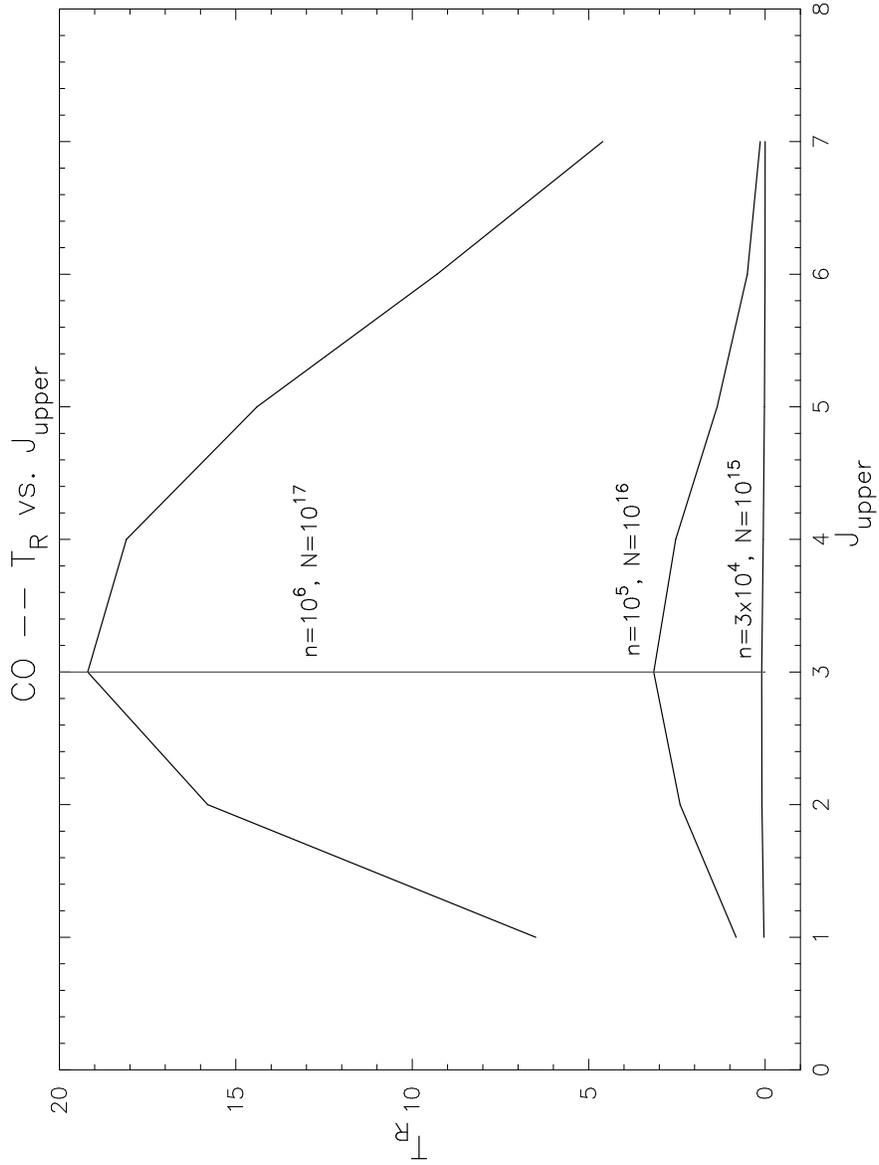}
\figcaption[fig2.ps]{Plot of radiation temperature versus lower J transition
over a range of physical conditions.  These LVG models indicate that
the J=3$\rightarrow$2 line is the peak over the range of values
relevant to most planetary nebulae.
\label{fig2}}
\end{figure}

\begin{figure}
\epsscale{0.7}
\plotone{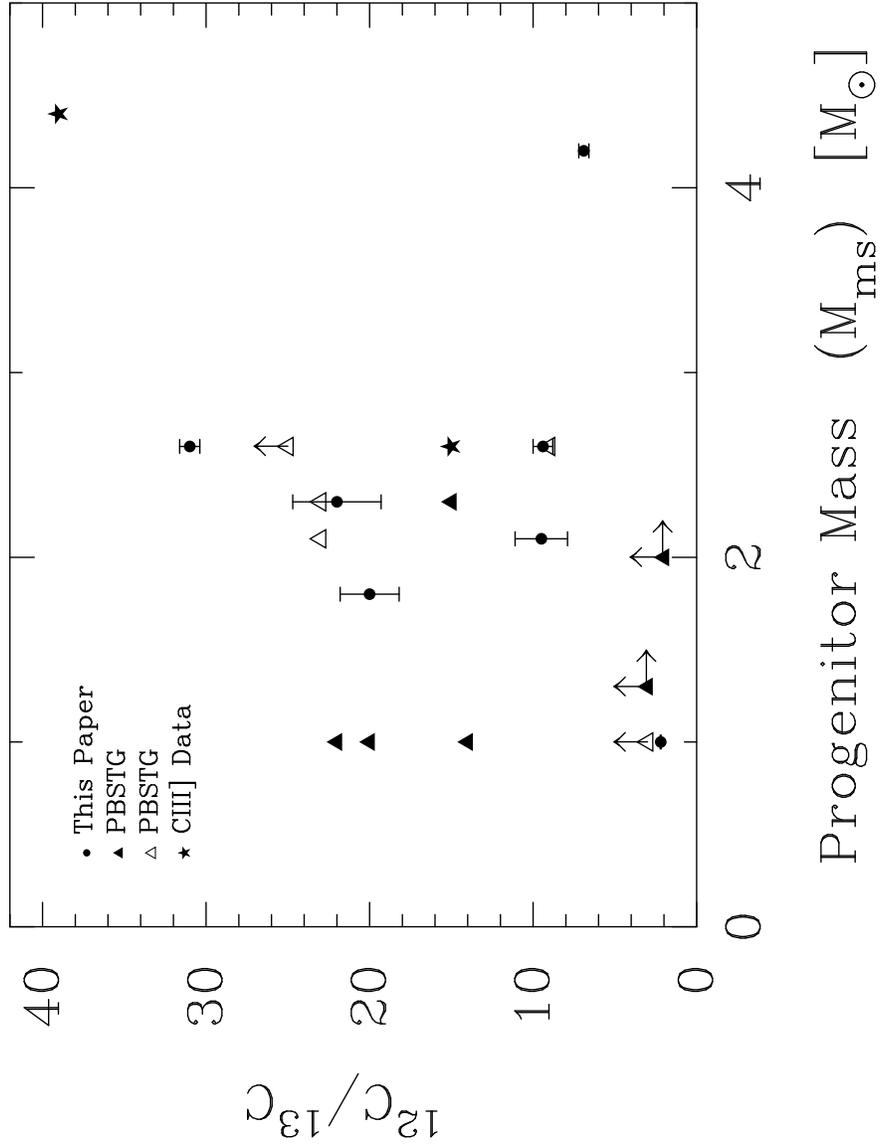}
\figcaption[fig3.ps]{The \ratc\ abundance ratios versus the progenitor
main sequence mass.  The circles are taken from this paper,
the triangles are from PBSTG, and the stars are taken from UV data.
The open triangles are from PBSTG and are in common with our sample.
\label{fig3}}
\end{figure}

\newpage

%
%
\begin{deluxetable}{lcccccccc}
\rotate
\tablewidth{0pt}
\tablecaption{Observed Planetary Nebulae}
\tablehead{
\colhead{} & \colhead{(${\ell},b$)} & \colhead{$\alpha$} &
\colhead{$\delta$} & \colhead{Offset} & \colhead{$V_{\rm LSR}$} &
\colhead{Size} & \colhead{$M_{\rm ms}$} & \colhead{}  \\
\colhead{Name} & \colhead{(degrees)} &  \colhead{(1950)} &
\colhead{(1950)} & \colhead{(arcsec)} & \colhead{(\ks)} & 
\colhead{(arcsec)} & \colhead{(\msun)} & \colhead{Shape$^{\rm a}$} }
\startdata
AFGL 618   & (166,$-$06) & 04 39 33.8 & $+$36 01 15 & ($0,0$) &
$-$22  & 12 & \nodata & E \\
M1-7       & (189,$-$07) & 06 34 18.0 & $+$24 03 00 & ($0,0$) &
$-$11  & 11 & 1.8 & E \\
NGC 2346   & (215,$+$03) & 07 06 49.6 & $-$00 43 30 & ($0,0$) &
$+$9   & 52 & 2.3 & B \\
M1-16      & (226,$+$05) & 07 34 55.4 & $-$09 32 00 & ($0,0$) &
$+$50  &$<$4& 1.0 & B \\
VV 47      & (164,$+$31) & 07 54 00.0 & $+$53 33 00 & ($-66,+66$) &
$-$55  &380 & \nodata & E \\
NGC 6072   & (342,$+$10) & 16 09 41.0 & $-$36 06 10 & ($0,0$) &
$+$15  & 70 & 4.2 & B \\
M4-9       & (024,$+$05) & 18 11 37.4 & $-$05 00 17 & ($0,+20$) &
$-$15  & 47 & \nodata & \nodata \\
NGC 6720   & (063,$+$13) & 18 51 43.7 & $+$32 57 56 & ($-40,-20$) &
$-$2.6 & 76 & 2.1 & E \\
NGC 7027   & (084,$-$03) & 21 05 09.4 & $+$42 02 03 & ($0,0$) &
$+$26  & 15 & 2.6 & E \\
IRAS 21282 &(094,$-$00)&21 28 15.1& $+$50 50 47 & ($0,0$) &
$+$18 & \nodata & \nodata & \nodata \\
NGC 7293   & (036,$-$57) & 22 26 54.8 & $-$21 05 41 & ($-372,0$) &
$-$24 & 660 & 2.6 & E \\

\enddata
\tablenotetext{a}{Planetary nebulae morphology: R = Round, E = Elliptical, and B = Bipolar.}
\end{deluxetable}
%
%
\begin{deluxetable}{llccc}
\tablewidth{0pt}
\tablecaption{Observing Log}
\tablehead{

\colhead{} & \colhead{} &
\colhead{Frequency} & \colhead{$\Theta_{\rm beam}$} &
\colhead{$\Delta v$} \\

\colhead{Telescope} & \colhead{Transition} &
\colhead{(GHz)} & \colhead{(arcsec)} &
\colhead{(\ks)} }
\startdata

NRAO 12\m\ & ($^{13}$CO) J=2$\rightarrow$1 & 220.399 & 29 & 1.4 \\
NRAO 12\m\ & ($^{12}$CO) J=2$\rightarrow$1        & 230.538 & 27 & 1.3 \\
HHT  10\m\ & ($^{13}$CO) J=3$\rightarrow$2 & 330.588 & 23 & 0.9 \\
HHT  10\m\ & ($^{12}$CO) J=3$\rightarrow$2        & 345.796 & 22 & 0.9 \\

\enddata
\end{deluxetable}
%
%
\begin{deluxetable}{lcccccc}              
\tablewidth{0pt}
\tablecaption{Results}
\tablehead{
\colhead{Source}      & \colhead{Transition} &
\colhead{$T_{\rm peak}$} & \colhead{$RMS$} &
\colhead{$V_{\rm centroid}$} & \colhead{$\Delta V^{\rm a}$}  & \colhead{$I$}\\
\colhead{}          &  \colhead{}       &
\colhead{(K)} & \colhead{(K)}
& \colhead{(\ks)} & \colhead{(\ks)} &\colhead{(K \ks)} }
\startdata
             & $^{12}$CO J=2$\rightarrow$1 & 2.4   & 0.09  & --22.0  & 23.9     & 60. $\pm$ 0.8   \\
AFGL 618     & $^{13}$CO J=2$\rightarrow$1 & 0.35  & 0.03  & --20.9  & 25.5     & 8.8  $\pm$  0.2  \\
             & $^{12}$CO J=3$\rightarrow$2 & 5.3   & 0.51  & --21.7  & 22.4     & 120.  $\pm$   4.0\\
             & $^{13}$CO J=3$\rightarrow$2 & 0.95  & 0.12  & --20.7  & 27.5     & 23.  $\pm$ 0.9  \\
             &                             &       &       &         &          &        \\
M1-7         & $^{12}$CO J=2$\rightarrow$1 & 0.70  & 0.02  & --10.7  & 35.1     & 22. $\pm$ 0.2    \\
             & $^{13}$CO J=2$\rightarrow$1 & 0.04  & 0.01  & --10.0  & 53.9     & 1.1  $\pm$ 0.1   \\
             & $^{12}$CO J=3$\rightarrow$2 & 1.5   & 0.06  & --11.5  & 39.0     & 49.   $\pm$ 0.5  \\
             & $^{13}$CO J=3$\rightarrow$2 & 0.17  & 0.06  & --4.2   & 45.0     & 3.0   $\pm$ 0.5  \\
             &                             &       &       &         &          &        \\
NGC 2346~~   & $^{12}$CO J=2$\rightarrow$1 & 1.1   & 0.02  & 6.0     & 10.1$^{\rm b}$ & 18.  $\pm$ 0.2\\
             & $^{13}$CO J=2$\rightarrow$1 & 0.04  & 0.01  & 6.6     & 31.8$^{\rm b}$ & 0.81  $\pm$ 0.1\\
             & $^{12}$CO J=3$\rightarrow$2 & 1.8   & 0.13  & 2.3    & 13.0     & 30.   $\pm$ 1.3  \\
             & $^{13}$CO J=3$\rightarrow$2 & $<$0.24  & 0.08  & \nodata & \nodata      & \nodata  \\
             &                             &       &       &         &          &        \\
             & $^{12}$CO J=2$\rightarrow$1 & 0.87  & 0.02  & 50.2    & 25.9     & 25.  $\pm$ 0.2   \\
M1-16        & $^{13}$CO J=2$\rightarrow$1 & 0.32  & 0.01  & 50.0    & 28.0     & 9.8  $\pm$ 0.1   \\
             & $^{12}$CO J=3$\rightarrow$2 & 1.8   & 0.10  & 49.4    & 26.2     & 49.   $\pm$ 0.8  \\
             & $^{13}$CO J=3$\rightarrow$2 & 0.80  & 0.11  & 48.6    & 31.1     & 21.   $\pm$ 1.0  \\
             &                             &       &       &         &          &        \\
VV 47        & $^{12}$CO J=2$\rightarrow$1 & 0.17  & 0.02  & --68.3  & 3.25     & 0.92  $\pm$ 0.1\\
             & $^{13}$CO J=2$\rightarrow$1 & $<$0.03&0.01  & \nodata     & \nodata      & \nodata \\
             &                             &       &       &         &          &        \\
NGC 6072     & $^{12}$CO J=2$\rightarrow$1 & 1.0   & 0.05  & 12.3    & 10.7$^{\rm b}$ & 18.   $\pm$ 0.4 \\
             & $^{13}$CO J=2$\rightarrow$1 & 0.24  & 0.01  & 11.5    & 5.16$^{\rm b}$ & 2.6   $\pm$ 0.1 \\
             &                             &       &       &         &          &        \\
M4-9         & $^{12}$CO J=2$\rightarrow$1 & 0.12  & 0.01  & --19.5  & 26.8     & 2.8  $\pm$ 0.1   \\
             & $^{13}$CO J=2$\rightarrow$1 & $<$0.03  & 0.01  & \nodata  & \nodata     & \nodata  \\
             &                             &       &       &         &          &        \\
             & $^{12}$CO J=2$\rightarrow$1 & 0.54  & 0.02  & --4.0   & 27.0     & 11.   $\pm$ 0.4  \\
NGC 6720     & $^{13}$CO J=2$\rightarrow$1 & 0.05  & 0.01  & --3.5   & 49.1     & 1.2   $\pm$ 0.2 \\
             & $^{12}$CO J=3$\rightarrow$2 & 1.4   & 0.11  & --4.7   & 16.3     & 22.   $\pm$ 0.9 \\
             & $^{13}$CO J=3$\rightarrow$2 & $<$0.12& 0.04 & \nodata     & \nodata      & \nodata \\
             &                             &       &       &         &          &        \\
             & $^{12}$CO J=2$\rightarrow$1 & 8.4   & 0.003 & 26.8    & 22.3     & 200. $\pm$ 2.0    \\
NGC 7027     & $^{13}$CO J=2$\rightarrow$1 & 0.37  & 0.004 & 22.8    & 27.4     & 6.4  $\pm$ 0.9   \\
             & $^{12}$CO J=3$\rightarrow$2 & 13.   & 0.65  & 26.     & 22.      & 300. $\pm$ 2.0  \\
             & $^{13}$CO J=3$\rightarrow$2 & $<$1.0&0.34  & \nodata     & \nodata      & \nodata \\
             &                             &       &       &         &          &        \\
             & $^{12}$CO J=2$\rightarrow$1 & 2.5   & 0.06  & 18.1    & 18.4     & 50.  $\pm$ 0.4  \\
IRAS 21282   & $^{13}$CO J=2$\rightarrow$1 & 0.06  & 0.01  & 15.6    & 18.8     & 0.66  $\pm$ 0.1  \\
             & $^{12}$CO J=3$\rightarrow$2 & 5.1   & 0.13  & 18.9    & 17.3     & 100.  $\pm$ 0.9 \\
             & $^{13}$CO J=3$\rightarrow$2 & 0.15  & 0.04  & 21.8    & 19.8     & 1.5   $\pm$ 0.3  \\
             &                             &       &       &         &          &        \\
             & $^{12}$CO J=2$\rightarrow$1 & 3.3   & 0.05  & --15.4  & 3.5      & 15.  $\pm$ 0.2 \\
NGC 7293     & $^{13}$CO J=2$\rightarrow$1 & 0.34  & 0.02  & --15.3  & 3.2      & 1.6  $\pm$ 0.1 \\
             & $^{12}$CO J=3$\rightarrow$2 & 0.97  & 0.12  & --14.8  & 2.6      & 2.5   $\pm$ 0.2\\
             & $^{13}$CO J=3$\rightarrow$2 & 0.29  & 0.08  & --15.7  & 2.3      & 0.52  $\pm$ 0.2\\
\enddata
\tablenotetext{a}{Based on a one-component Gaussian model.}
\tablenotetext{b}{Multiple components exists within the single Gaussian fit.}
\end{deluxetable}
%
%
\begin{deluxetable}{lccccc}              
\tablewidth{0pt}
\tablecaption{\ratc\ Abundance Ratios}
\tablehead{
\colhead{} & \colhead{} & 
\multicolumn{4}{c}{\underline{~~~~~~~~~~~~~~~~~~~~~~~~~~\ratc\
Ratio~~~~~~~~~~~~~~~~~~~~~~~~~~~~}} \\
\colhead{} & \colhead{} & \colhead{NRAO 12\m} & 
\colhead{HHT 10\m} & \colhead{} & \colhead{} \\
\colhead{Source}      & \colhead{PK} &
\colhead{(J=2$\rightarrow$1)$^{\rm a}$} & 
\colhead{(J=3$\rightarrow$2)$^{\rm a}$} & \colhead{Adopted$^{\rm b}$} &
\colhead{PBSTG$^{\rm c}$}
}

\startdata

AFGL 618   & (166--06)  & 6.8 (6.9)      & 5.2 (5.5)  
& $>$ 4.6$\,^{\rm d}$   & 4$\,^{\rm e}$  \\

M1--7      & (189--07)  & 20. (18.)      & $>$6.7        
& 20. $\pm$ 1.8         & \nodata          \\

NGC 2346   & (215+03)   & 22. (28.)      & $>$6.3        
& 22.  $\pm$ 2.7        & 23           \\

M1--16     & (226+05)   & 2.6 (2.7)      & 2.4 (2.2)  
& 2.2 $\pm$ 0.03$\,^{\rm d}$  & 3$\,^{\rm e}$  \\

VV 47      & (164+31)   & $>$ 5.7        & \nodata        
& $>$ 5.7         & \nodata          \\

NGC 6072   & (342+10)   & 6.9 (4.2)      & \nodata        
& 6.9  $\pm$ 0.31            & \nodata          \\

M4--9      & (024+05)   & $>$ 4.0      & \nodata        
& $>$ 4.0             & 18           \\

NGC 6720   & (063+13)   & 9.5 (10.8)     & $>$ 14.    
& 9.5  $\pm$ 1.6            & 22           \\

NGC 7027   & (084--03)  & 31.(22.)       & $>$ 13.    
& 31.  $\pm$ 0.62            & 25$\,^{\rm e}$ \\

IRAS 21282 &  \nodata       & 76. (42.)      & 66. (33.)  
& $>$ 32.$\,^{\rm d}$   & \nodata          \\

NGC 7293   & (036--57)  & 9.4 (9.7)      & 4.8 (3.3)  
& 9.4  $\pm$ 0.60           & 9            \\

\enddata
\tablenotetext{a}{Based on CO integrated intensities and 
peak intensity (in parentheses).}
\tablenotetext{b}{Adopted \ratc\ ratio (see text).}
\tablenotetext{c}{From Palla et al. (2000).}
\tablenotetext{d}{Based on LVG model.}
\tablenotetext{e}{Lower limit based on optically thick CO.}
\end{deluxetable}

\end{document}